\documentclass[aps,prl,twocolumn,showpacs,superscriptaddress,preprintnumbers,floatfix]{revtex4}
\usepackage{graphicx}
\usepackage{amsmath}

\def\vermark{}
\def\bsjetaBF{\ensuremath{[5.10\pm 0.50(\mathrm{stat.})\pm 0.25(\mathrm{syst.})
^{+1.14}_{-0.79}(N_{B_s^{(*)}\bar B_s^{(*)}})]\times 10^{-4}}}
\def\bsjetapBF{\ensuremath{[3.71\pm 0.61(\mathrm{stat.})\pm 0.18(\mathrm{syst.})
^{+0.83}_{-0.57}(N_{B_s^{(*)}\bar B_s^{(*)}})]\times 10^{-4}}}
\def\bsrBF{\ensuremath{0.73\pm 0.14(\mathrm{stat.})\pm 0.02(\mathrm{syst.})}}

\begin{document}
\preprint{
  Belle Preprint 2012-4
}
\preprint{
  KEK Preprint 2011-26
}

\makeatletter
\newcommand{\ps@mytitle}{%
  \renewcommand{\@oddhead}{{\vermark}\hfil}%
}
\makeatother

\title{First observation of $B_s^0\to J/\psi\eta$ and
$B_s^0\to J/\psi\eta'$}
\affiliation{University of Bonn, Bonn}
\affiliation{Budker Institute of Nuclear Physics SB RAS and Novosibirsk State University, Novosibirsk 630090}
\affiliation{Faculty of Mathematics and Physics, Charles University, Prague}
\affiliation{University of Cincinnati, Cincinnati, Ohio 45221}
\affiliation{Department of Physics, Fu Jen Catholic University, Taipei}
\affiliation{Justus-Liebig-Universit\"at Gie\ss{}en, Gie\ss{}en}
\affiliation{Gifu University, Gifu}
\affiliation{Gyeongsang National University, Chinju}
\affiliation{Hanyang University, Seoul}
\affiliation{University of Hawaii, Honolulu, Hawaii 96822}
\affiliation{High Energy Accelerator Research Organization (KEK), Tsukuba}
\affiliation{Indian Institute of Technology Guwahati, Guwahati}
\affiliation{Indian Institute of Technology Madras, Madras}
\affiliation{Indiana University, Bloomington, Indiana 47408}
\affiliation{Institute of High Energy Physics, Chinese Academy of Sciences, Beijing}
\affiliation{Institute of High Energy Physics, Vienna}
\affiliation{Institute of High Energy Physics, Protvino}
\affiliation{Institute for Theoretical and Experimental Physics, Moscow}
\affiliation{J. Stefan Institute, Ljubljana}
\affiliation{Kanagawa University, Yokohama}
\affiliation{Institut f\"ur Experimentelle Kernphysik, Karlsruher Institut f\"ur Technologie, Karlsruhe}
\affiliation{Korea Institute of Science and Technology Information, Daejeon}
\affiliation{Korea University, Seoul}
\affiliation{Kyungpook National University, Taegu}
\affiliation{\'Ecole Polytechnique F\'ed\'erale de Lausanne (EPFL), Lausanne}
\affiliation{Faculty of Mathematics and Physics, University of Ljubljana, Ljubljana}
\affiliation{Luther College, Decorah, Iowa 52101}
\affiliation{University of Maribor, Maribor}
\affiliation{Max-Planck-Institut f\"ur Physik, M\"unchen}
\affiliation{University of Melbourne, School of Physics, Victoria 3010}
\affiliation{Graduate School of Science, Nagoya University, Nagoya}
\affiliation{Kobayashi-Maskawa Institute, Nagoya University, Nagoya}
\affiliation{Nara Women's University, Nara}
\affiliation{National Central University, Chung-li}
\affiliation{National United University, Miao Li}
\affiliation{Department of Physics, National Taiwan University, Taipei}
\affiliation{H. Niewodniczanski Institute of Nuclear Physics, Krakow}
\affiliation{Nippon Dental University, Niigata}
\affiliation{Niigata University, Niigata}
\affiliation{University of Nova Gorica, Nova Gorica}
\affiliation{Osaka City University, Osaka}
\affiliation{Pacific Northwest National Laboratory, Richland, Washington 99352}
\affiliation{Panjab University, Chandigarh}
\affiliation{Research Center for Nuclear Physics, Osaka University, Osaka}
\affiliation{University of Science and Technology of China, Hefei}
\affiliation{Seoul National University, Seoul}
\affiliation{Sungkyunkwan University, Suwon}
\affiliation{School of Physics, University of Sydney, NSW 2006}
\affiliation{Tata Institute of Fundamental Research, Mumbai}
\affiliation{Excellence Cluster Universe, Technische Universit\"at M\"unchen, Garching}
\affiliation{Toho University, Funabashi}
\affiliation{Tohoku Gakuin University, Tagajo}
\affiliation{Tohoku University, Sendai}
\affiliation{Department of Physics, University of Tokyo, Tokyo}
\affiliation{Tokyo Institute of Technology, Tokyo}
\affiliation{Tokyo Metropolitan University, Tokyo}
\affiliation{Tokyo University of Agriculture and Technology, Tokyo}
\affiliation{CNP, Virginia Polytechnic Institute and State University, Blacksburg, Virginia 24061}
\affiliation{Yamagata University, Yamagata}
\affiliation{Yonsei University, Seoul}
  \author{J.~Li}\affiliation{Seoul National University, Seoul} 
  \author{I.~Adachi}\affiliation{High Energy Accelerator Research Organization (KEK), Tsukuba} 
  \author{H.~Aihara}\affiliation{Department of Physics, University of Tokyo, Tokyo} 
  \author{K.~Arinstein}\affiliation{Budker Institute of Nuclear Physics SB RAS and Novosibirsk State University, Novosibirsk 630090} 
  \author{D.~M.~Asner}\affiliation{Pacific Northwest National Laboratory, Richland, Washington 99352} 
  \author{V.~Aulchenko}\affiliation{Budker Institute of Nuclear Physics SB RAS and Novosibirsk State University, Novosibirsk 630090} 
  \author{T.~Aushev}\affiliation{Institute for Theoretical and Experimental Physics, Moscow} 
  \author{A.~M.~Bakich}\affiliation{School of Physics, University of Sydney, NSW 2006} 
  \author{V.~Bhardwaj}\affiliation{Nara Women's University, Nara} 
  \author{B.~Bhuyan}\affiliation{Indian Institute of Technology Guwahati, Guwahati} 
% ...
\collaboration{The Belle Collaboration}

\begin{abstract}
We report first observations of $B_s^0\to J/\psi\eta$ and 
$B_s^0\to J/\psi\eta'$.  The results are obtained from
$121.4\;\mathrm{fb}^{-1}$
of data collected at the $\Upsilon(5S)$ resonance with the Belle detector at
the KEKB $e^+e^-$ collider.  We obtain the branching fractions
$\mathcal{B}(B_s^0\to J/\psi\eta)=\bsjetaBF$, and
$\mathcal{B}(B_s^0\to J/\psi\eta')=\bsjetapBF$.  The ratio of the two branching
fractions is measured to be 
$\frac{\mathcal{B}(B_s\to J/\psi\eta')}{\mathcal{B}(B_s\to J/\psi \eta)} = \bsrBF$.
\end{abstract}
\pacs{13.25.Hw, 14.40.Nd, 13.25.Gv, 14.40.Be}

\maketitle
\thispagestyle{mytitle}
\markright{\vermark}

\begin{figure}
\includegraphics[width=\columnwidth]
{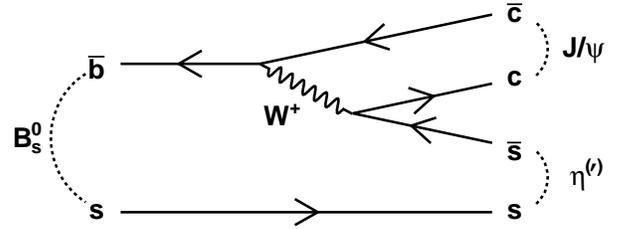}
\caption{\label{fig:bs_fyman}
Dominant diagram for the processes $B_s^0\to J/\psi\eta^{(\prime)}$.}
\end{figure}

The decays $B_s^0\to J/\psi \eta^{(\prime)}$ are dominated by the
$b\to c\bar c s$ process shown in Fig.~\ref{fig:bs_fyman}.  The
$J/\psi \eta^{(\prime)}$ final states 
are $CP$-even eigenstates; their time distributions can be used
to directly measure the $B_s^0$ width difference
$\Delta\Gamma_s$ and the $CP$-violating phase $\phi_s$~\cite{Dunietz:2000cr}
without an angular analysis.
Assuming flavor SU(3) symmetry and factorization, the $B_s^0\to J/\psi \eta^{(\prime)}$
branching fractions relative to the decay $B_d^0\to J/\psi K^0$
are estimated to be~\cite{Skands:2000ru}:
\begin{equation*}
\frac{
\mathcal{B}(B_s^0\to J/\psi\eta^{(\prime)})}
{\mathcal{B}(B_d^0\to J/\psi K^0)} =
\sin^2\phi_P(\cos^2\phi_P)\times p_{B_s^0}^{*3}/p_{B_d^0}^{*3},
\end{equation*}
where $p^*$ is the momentum of
$J/\psi$ in the rest frame of the $B_s^0$ or
$B_d^0$.  Here $\phi_P= (41.4\pm 0.5)^\circ$~\cite{Ambrosino:2009sc}
is the pseudoscalar mixing angle in
the flavor basis with $\eta(\eta') = \frac{1}{\sqrt{2}}[u\bar u+d \bar d]
\cos\phi_P(\sin\phi_P) -(+) s\bar s\sin\phi_P(\cos\phi_P)$, 
and other possible flavor singlet content of the $\eta'$
such as gluonium is neglected.  Using this relation and the value
$\mathcal{B}(B_d^0\to J/\psi K^0) = 8.71\times
10^{-4}$~\cite{Nakamura:2010zzi}, we expect $\mathcal{B}(B_s^0\to
J/\psi\eta^{(\prime)})\sim 4.16(4.31)\times 10^{-4}$.
The ratio of the two branching fractions $\mathcal{B}(B_s^0\to
J/\psi\eta')/\mathcal{B}(B_s^0\to J/\psi\eta)$
is expected to be $1.04\pm 0.04$.
This ratio estimation does not require flavor SU(3) or the assumption
of factorization~\cite{Fleischer:2011ib} and can be used to test the
$\eta-\eta'$ mixing scheme~\cite{Thomas:2007uy,Fleischer:2011ib}.
The only previous experimental result for these decay channels
is the $90\%$ confidence level upper limit
$\mathcal{B}(B_s^0\to J/\psi\eta) < 3.8\times
10^{-3}$~\cite{Acciarri:1996ur}. 

In this Letter, we report measurements of fully reconstructed
$B_s^0\to J/\psi\eta$ and $B_s^0\to J/\psi\eta'$ decays using a
$ 121.4\;\mathrm{fb}^{-1}$ data sample collected with the Belle
detector at the KEKB asymmetric-energy $e^+e^-$ collider~\cite{KEKB}
operated at the $\Upsilon(5S)$ resonance.  $B_s^0$ mesons can be
produced in three $\Upsilon(5S)$ decays: $\Upsilon(5S) \to B_s^*
\bar{B}_s^*$, $B_s^*\bar{B}_s^0$, and $B_s^0 \bar{B}_s^0$ where the
$B_s^*$ mesons decay to $B_s^0 \gamma$.  The number of $B_s^{(*)}\bar
B_s^{(*)}$ pairs in the sample is measured to be $N_{B_s^{(*)}\bar
B_s^{(*)}} = (7.1 \pm 1.3) \times 10^6$ using inclusive $D_s$ production
methods described in Refs.~\cite{cleo-fs,belle-fs}. 
The fractions of $B_s^0$ production channels are defined as
$f_{B_s^*\bar B_s^*}=N_{B_s^*\bar B_s^*}/N_{B_s^{(*)}\bar B_s^{(*)}}$,
$f_{B_s^*\bar B_s^0}=N_{B_s^*\bar B_s^0}/N_{B_s^{(*)}\bar B_s^{(*)}}$.

The Belle detector is a large-solid-angle magnetic spectrometer
that consists of a silicon vertex detector, a 50-layer central drift
chamber (CDC), an array of aerogel threshold Cherenkov counters (ACC), a
barrel-like arrangement of time-of-flight scintillation counters, and
an electromagnetic calorimeter (ECL) comprised of CsI(Tl) crystals
located inside a superconducting solenoid coil that provides a 1.5 T magnetic field.
An iron flux-return located outside the coil is instrumented to detect
$K_L^0$ mesons and identify muons.  The detector is described in detail
elsewhere~\cite{:2000cg}.

Charged tracks are required to originate within $0.5$ cm in the
radial direction and within 5 cm in the beam direction,
with respect to the interaction point.  Electron candidates
are identified by combining information from the ECL, the CDC $(dE/dx)$, and
the ACC.  Muon candidates are identified through track penetration depth
and hit patterns in the $K_L^0$ muon system.  The identification of pions is based on
combining information from the CDC $(dE/dx)$, the time-of-flight scintillation
counters, and the ACC.

Pairs of  oppositely charged leptons $l^+l^-$ ($l=e\;\mathrm{or}\;\mu$) and
bremsstrahlung photons lying within 50 mrad of $e^+$ or $e^-$ tracks
are combined to form $J/\psi$ meson candidates.  The leptons are
required to be positively identified as electrons or muons and the dilepton
invariant mass is required to lie in the ranges
$-150\;\mathrm{MeV}/c^2<M_{ee(\gamma)}-
m_{J/\psi}<36\;\mathrm{MeV}/c^2$ and
$-60\;\mathrm{MeV}/c^2<M_{\mu\mu}- m_{J/\psi}<36\;\mathrm{MeV}/c^2$,
where $m_{J/\psi}$ denotes the nominal $J/\psi$ mass~\cite{Nakamura:2010zzi}, and
$M_{ee(\gamma)}$ and $M_{\mu\mu}$ are the reconstructed invariant masses
for $e^+e^-(\gamma)$ and $\mu^+\mu^-$, respectively.

Photon candidates are selected from ECL showers that are not associated with
charged tracks.  An energy deposition with a photonlike shower shape
and an energy greater than 50 MeV is required. 
Candidate $\pi^0\to\gamma\gamma$ decays are selected by combining two photon
candidates with an invariant mass in the range
$115\;\mathrm{MeV}/c^2<M_{\gamma\gamma} <155\;\mathrm{MeV}/c^2$ .

Candidate $\eta$ mesons are reconstructed in the $\gamma\gamma$ and
$\pi^+\pi^-\pi^0$ final states.  We require the invariant mass to be
in the range $500\;\mathrm{MeV}/c^2 < M_{\gamma\gamma} <
575\;\mathrm{MeV}/c^2\;([-3.5\sigma,2.0\sigma])$ and
$535\;\mathrm{MeV}/c^2 < M_{\pi^+\pi^-\pi^0} <
560\;\mathrm{MeV}/c^2\;(\pm 2.5\sigma)$. 

Candidate $\eta'$ mesons are reconstructed in the $\eta\pi^+\pi^-$ and
$\rho^0\gamma$ channels. Since $\eta$ candidates are
selected in two channels, there are three subchannels for
$\eta'$ reconstruction. Candidate $\rho^0\to\pi^+\pi^-$ decays
are oppositely charged pion pairs satisfying $550 \;
\mathrm{MeV}/c^2<M_{\pi^+\pi^-}<900\;\mathrm{MeV}/c^2$ and a helicity angle
requirement $|\cos\theta_\mathrm{hel}|<0.85$ since the $\rho^0$ in
$\eta' \to \rho^0\gamma$ is
longitudinally polarized.  Here $\theta_\mathrm{hel}$ is the helicity
angle of $\rho^0$, calculated as the angle between the direction of
the $\pi^+$ and the direction opposite to the $\eta'$ momentum in the
$\rho^0$ rest frame.  We require the reconstructed $\eta'$ invariant
mass to satisfy $ 940\;\mathrm{MeV}/c^2< M_{\eta'} <
975\;\mathrm{MeV}/c^2\;(\pm 3\sigma)$.

We combine $J/\psi$ and $\eta^{(\prime)}$ candidates to form $B_s^0$ mesons.
Signal candidates are identified by two kinematic variables computed
in the $\Upsilon(5S)$ rest frame: the energy difference $\Delta E=
E_B^*-E_\mathrm{beam}$ and the beam-energy constrained mass
$M_\mathrm{bc}=\sqrt{(E_\mathrm{beam})^2- (p_B^*)^2}$, where $E_B^*$
and $p_B^*$ are the energy and momentum of the reconstructed $B_s^0$
candidate.
To improve the $\Delta E$ and $M_\mathrm{bc}$ resolutions, mass-constrained
kinematic fits are applied to $J/\psi$, $\pi^0$, and $\eta^{(\prime)}$
candidates.  We retain $B_s^0$ meson candidates with $|\Delta E|<0.4$ GeV and
$M_\mathrm{bc}>5.25\;\mathrm{GeV}/c^2$ for further analysis.
The candidate that has a minimum sum of $\chi^2$'s for the
mass-constrained fits is selected if there is more than one
candidate.

The background is dominated by two-jet-like continuum events of the type
$e^+e^-\to q\bar q\;(q=u,d,s,c)$, together with other $B$ meson decay modes
($B=B_s^0,\;B_d^0,\;B^\pm$).
To suppress the continuum background, we require the ratio of second to zeroth
Fox-Wolfram moments~\cite{Fox:1978vu} to be less than $0.4$.  This
requirement is optimized by maximizing a figure of merit
$N_\mathrm{S}/\sqrt{N_\mathrm{S}+N_\mathrm{B}}$, where $N_\mathrm{S}$ is the
expected number of signal events and $N_\mathrm{B}$ is the number of background
events estimated from Monte Carlo simulation, in the $B_s^*\bar B_s^*$ signal region.

Signal and background distributions in $\Delta E$ and $M_\mathrm{bc}$ after
all selections are parametrized separately for each
$B_s^0\to J/\psi\eta^{(\prime)}$ subchannel.
The signal shapes for the two $\eta$ (three $\eta'$) subchannels are described with
a Crystal Ball function~\cite{crystalb:} 
(the sum of a Crystal Ball and a Gaussian function)
in $\Delta E$ and a Crystal Ball function in $M_\mathrm{bc}$. 
The means and widths of the distributions are calibrated with respect to
Monte Carlo values using a control sample of
$B^+\to J/\psi K^{*+}(K^{*+}\to K^+\pi^0)$ decays collected at 
the  $\Upsilon(4S)$ resonance.
The background shapes for all $\eta^{(\prime)}$ subchannels are smooth
and described with an exponential function in
$\Delta E$ and an ARGUS function~\cite{Albrecht:1986nr} in $M_\mathrm{bc}$.

An unbinned, extended maximum likelihood fit is performed simultaneously
to the total five two-dimensional $\Delta E-M_\mathrm{bc}$ distributions.
The branching fraction of each signal mode is a common parameter shared
among the corresponding $\eta^{(\prime)}$ subchannels. 
The parameters $f_{B_s^*\bar B_s^*}$ and $f_{B_s^*\bar B_s^0}$
are also common to all five subchannels.

\begin{figure}
\setlength{\unitlength}{0.005\columnwidth}
\mbox{\hspace{-0.02\columnwidth}%
\includegraphics[width=0.53\columnwidth]
{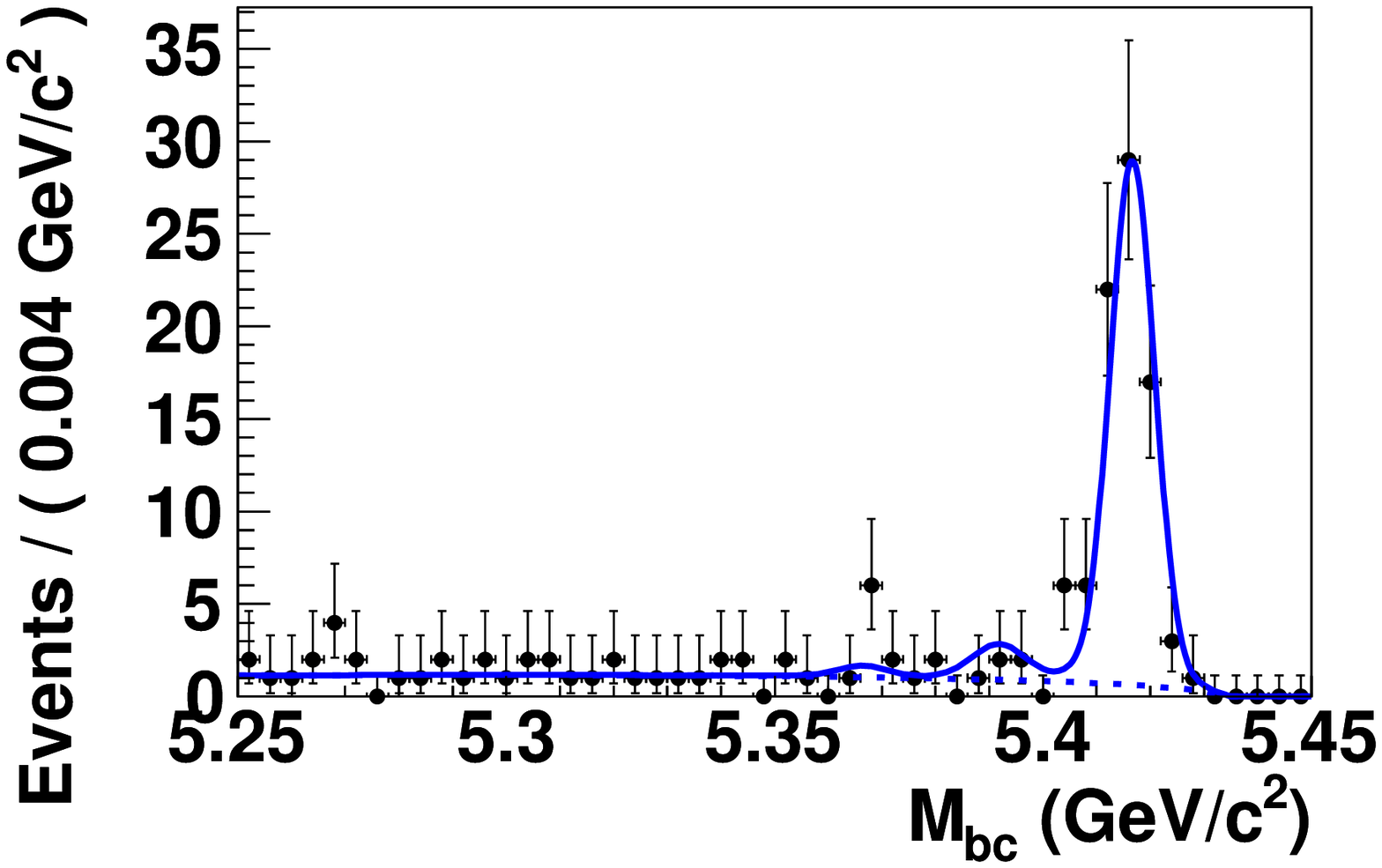}%
\hspace{-0.02\columnwidth}%
\includegraphics[width=0.53\columnwidth]
{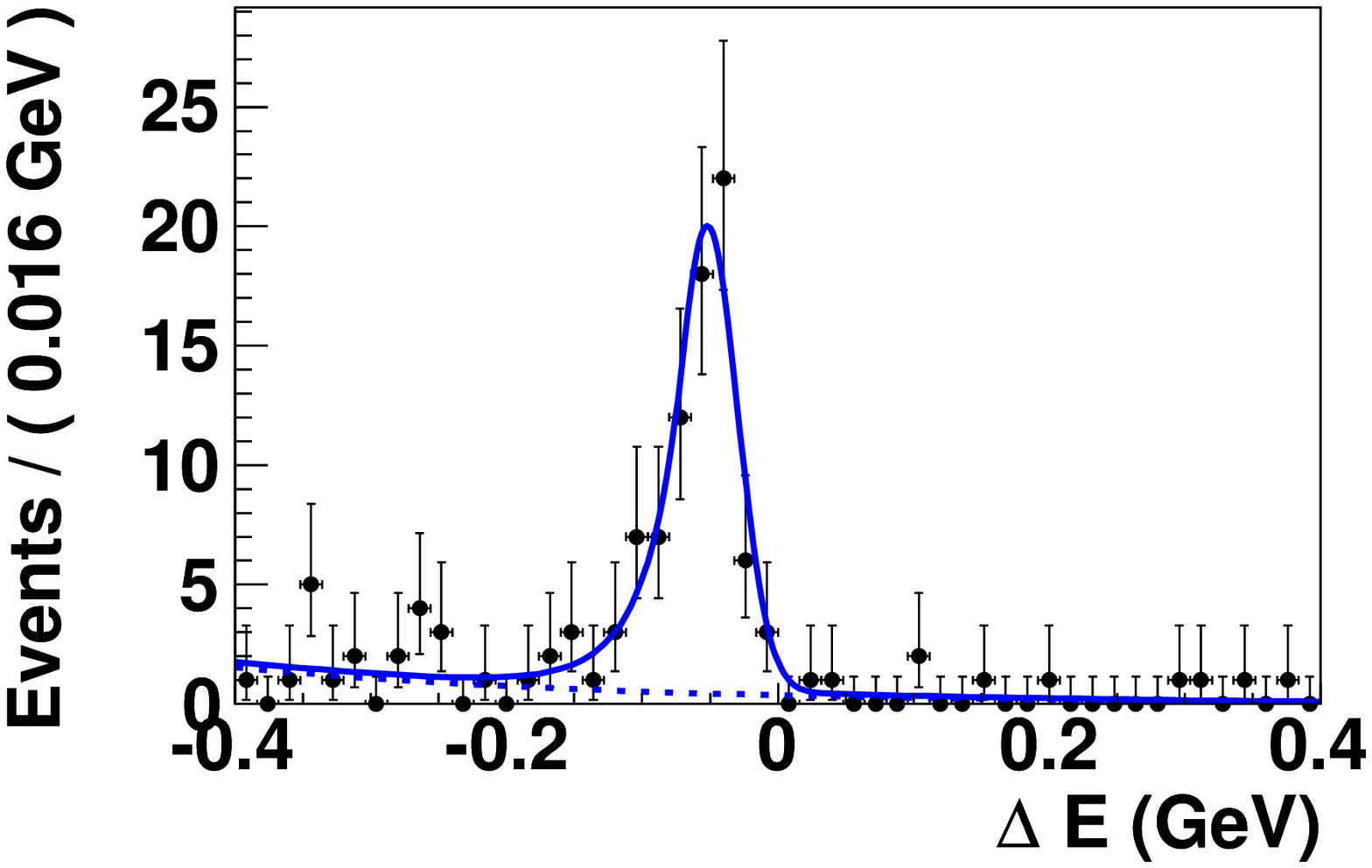}%
\hspace{-0.02\columnwidth}%
\hspace{-1\columnwidth}%
\begin{picture}(199,0)
\put(25,50){(a)}
\put(125,50){(b)}
\end{picture}\hfill}\\
\mbox{\hspace{-0.02\columnwidth}%
\includegraphics[width=0.53\columnwidth]
{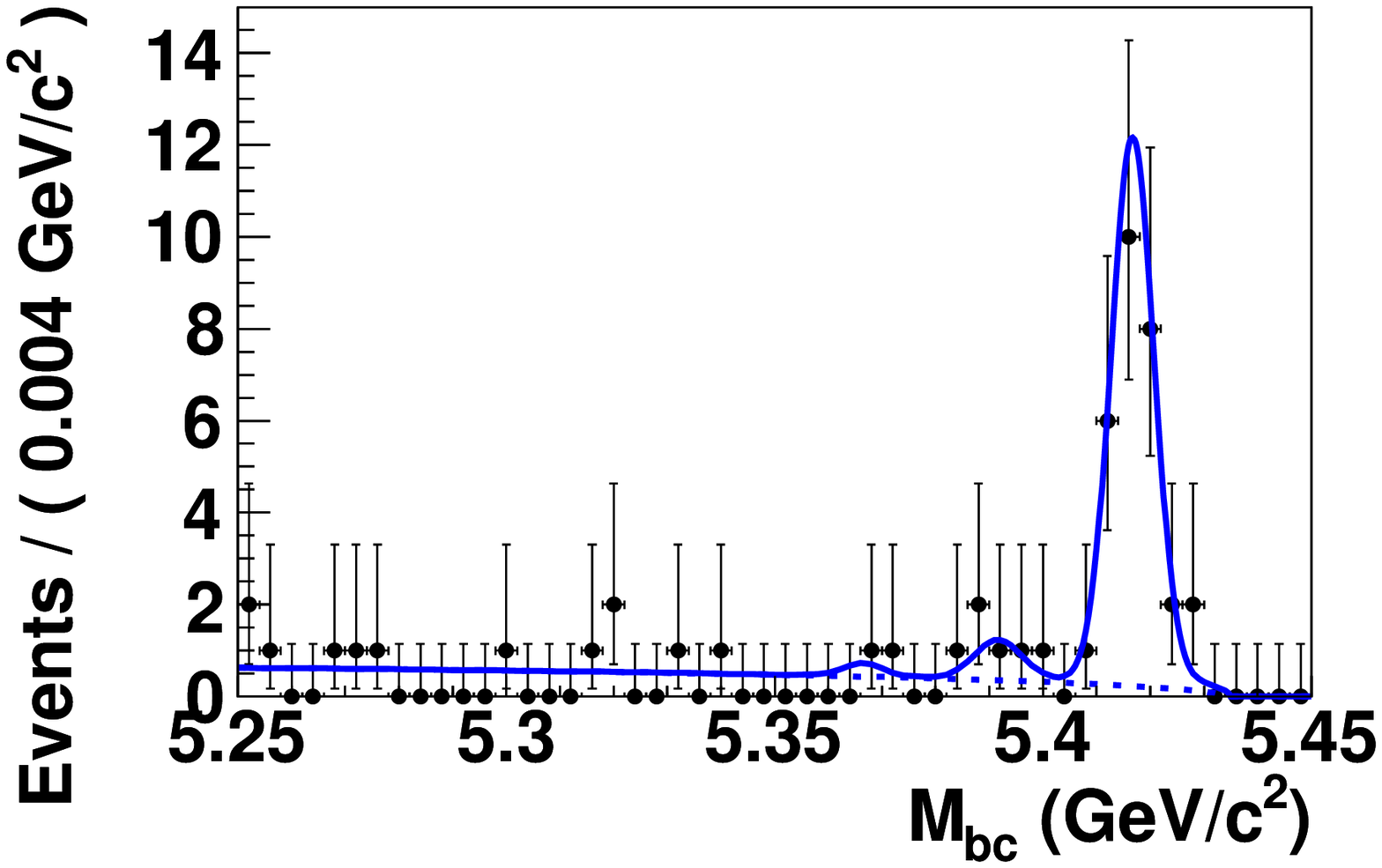}%
\hspace{-0.02\columnwidth}%
\includegraphics[width=0.53\columnwidth]
{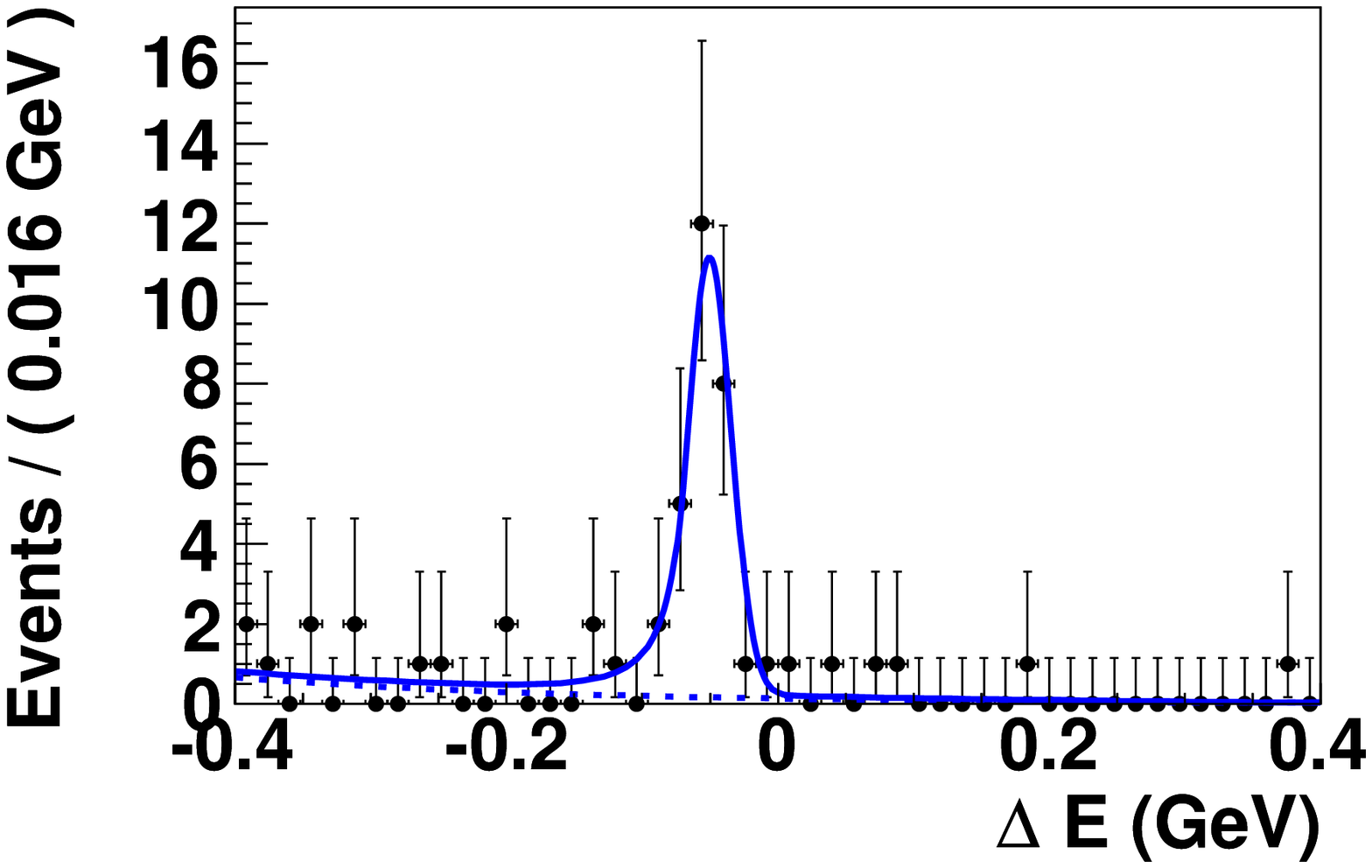}%
\hspace{-0.02\columnwidth}%
\hspace{-1\columnwidth}%
\begin{picture}(199,0)
\put(25,50){(c)}
\put(125,50){(d)}
\end{picture}\hfill}\\
\caption{\label{fig:5S_scan-eA-hde_mbc}
$M_\mathrm{bc}$ and $\Delta E$ distributions for the
$J/\psi\eta(\gamma\gamma)$ channel (a), (b) and 
the $J/\psi\eta(\pi^+\pi^-\pi^0)$ channel (c), (d). 
The projections are shown in the $B_s^*\bar B_s^*$ signal region with
$\Delta E \in [-116,12]$ MeV (a), (c), and with
$M_\mathrm{bc}\in [5.405,5.428]\;\mathrm{GeV}/c^2$ (b), (d).
Solid curves show projections of fit results.
Backgrounds are represented by the blue dotted curves.
Two small bumps around 5.37 and 5.39 GeV$/c^2$ in (a), (c) are
contributions from $B_s^0 \bar B_s^0$ and $B_s^*\bar B_s^0$ production channels,
due to the overlap of the $\Delta E$ signal regions. }
\end{figure}

\begin{figure}
\setlength{\unitlength}{0.005\columnwidth}
\mbox{\hspace{-0.02\columnwidth}%
\includegraphics[width=0.53\columnwidth]
{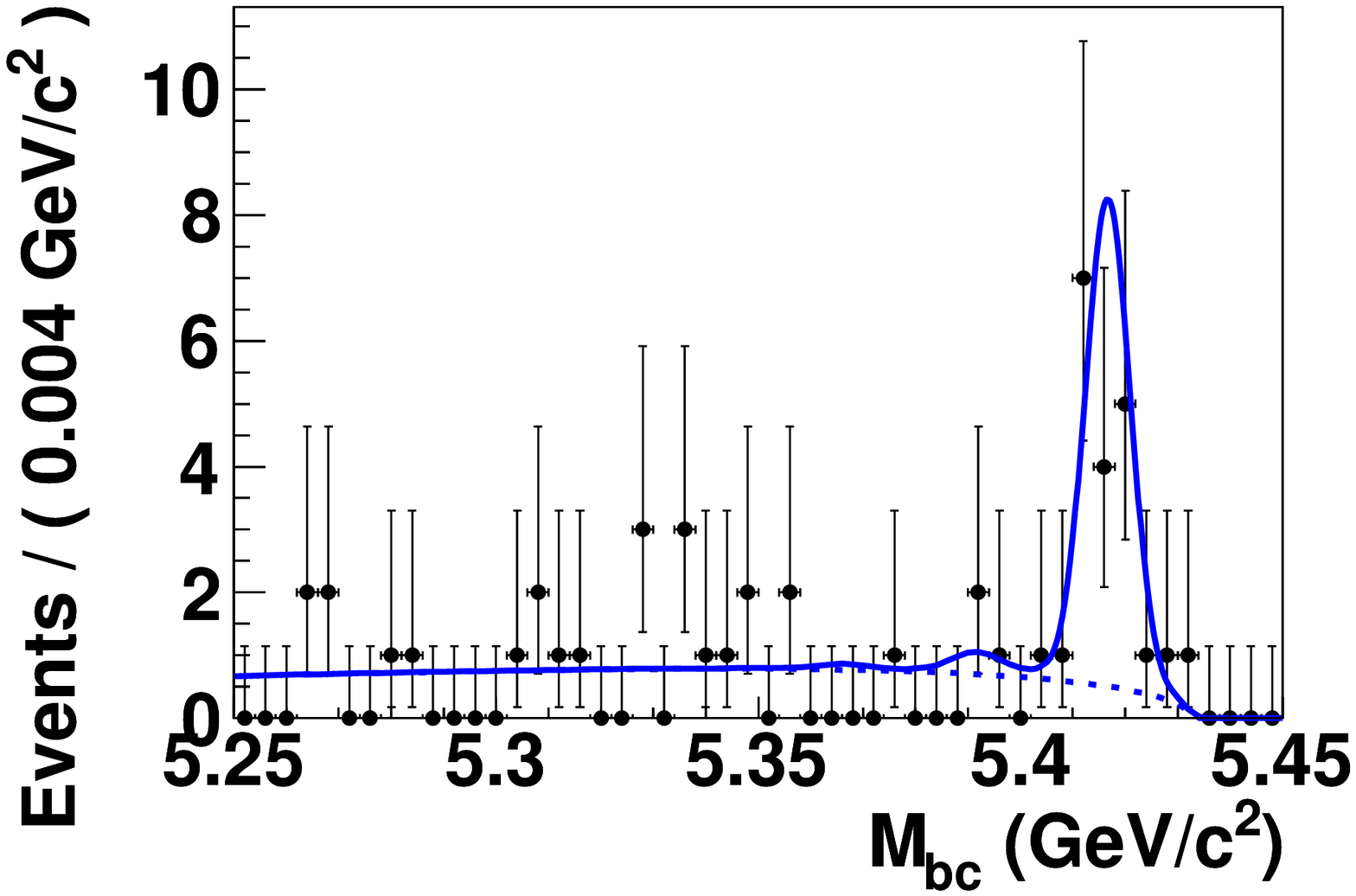}%
\hspace{-0.02\columnwidth}%
\includegraphics[width=0.53\columnwidth]
{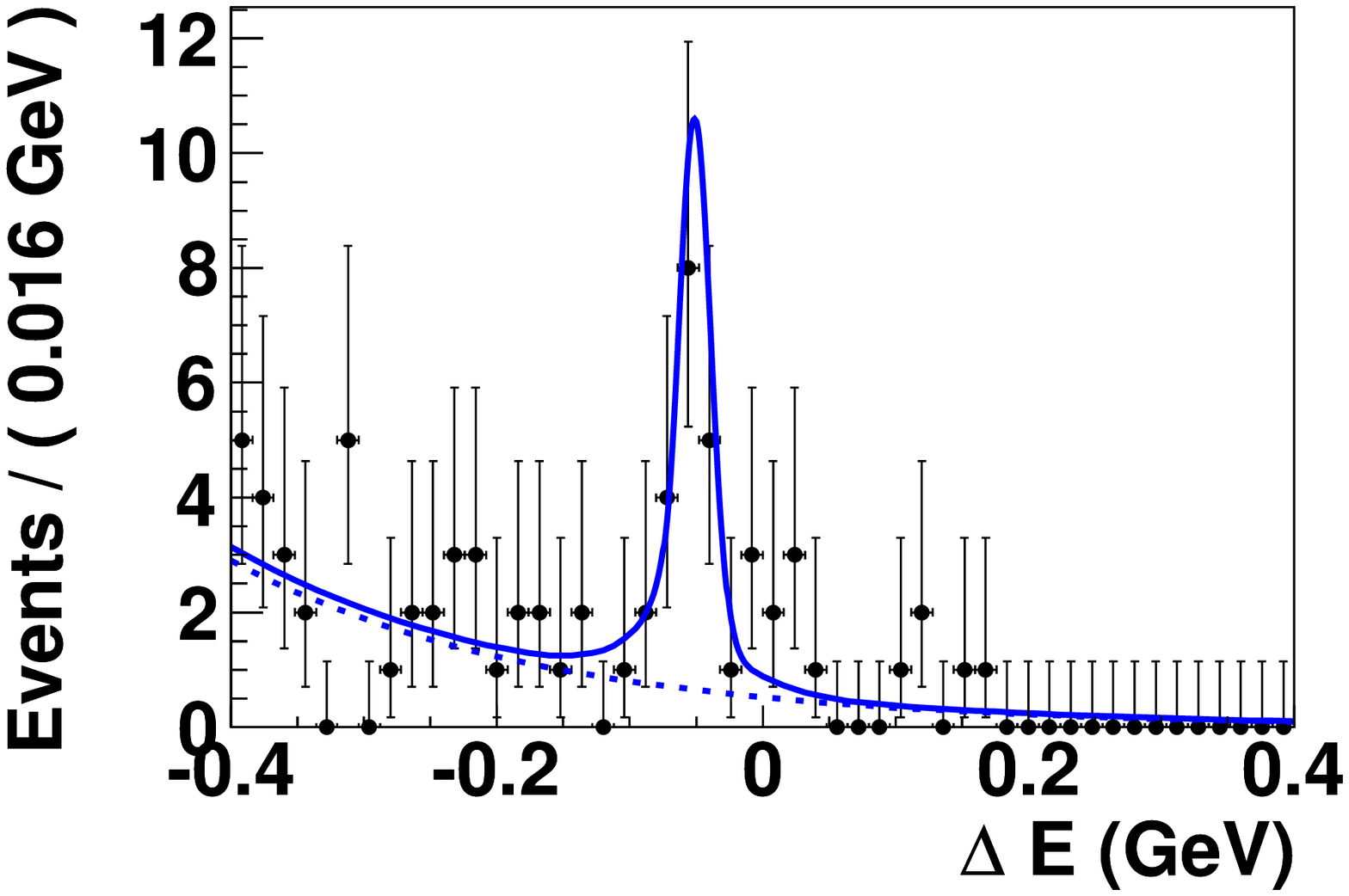}%
\hspace{-0.02\columnwidth}%
\hspace{-1\columnwidth}%
\begin{picture}(199,0)
\put(25,50){(a)}
\put(125,50){(b)}
\end{picture}\hfill}\\
\mbox{\hspace{-0.02\columnwidth}%
\includegraphics[width=0.53\columnwidth]
{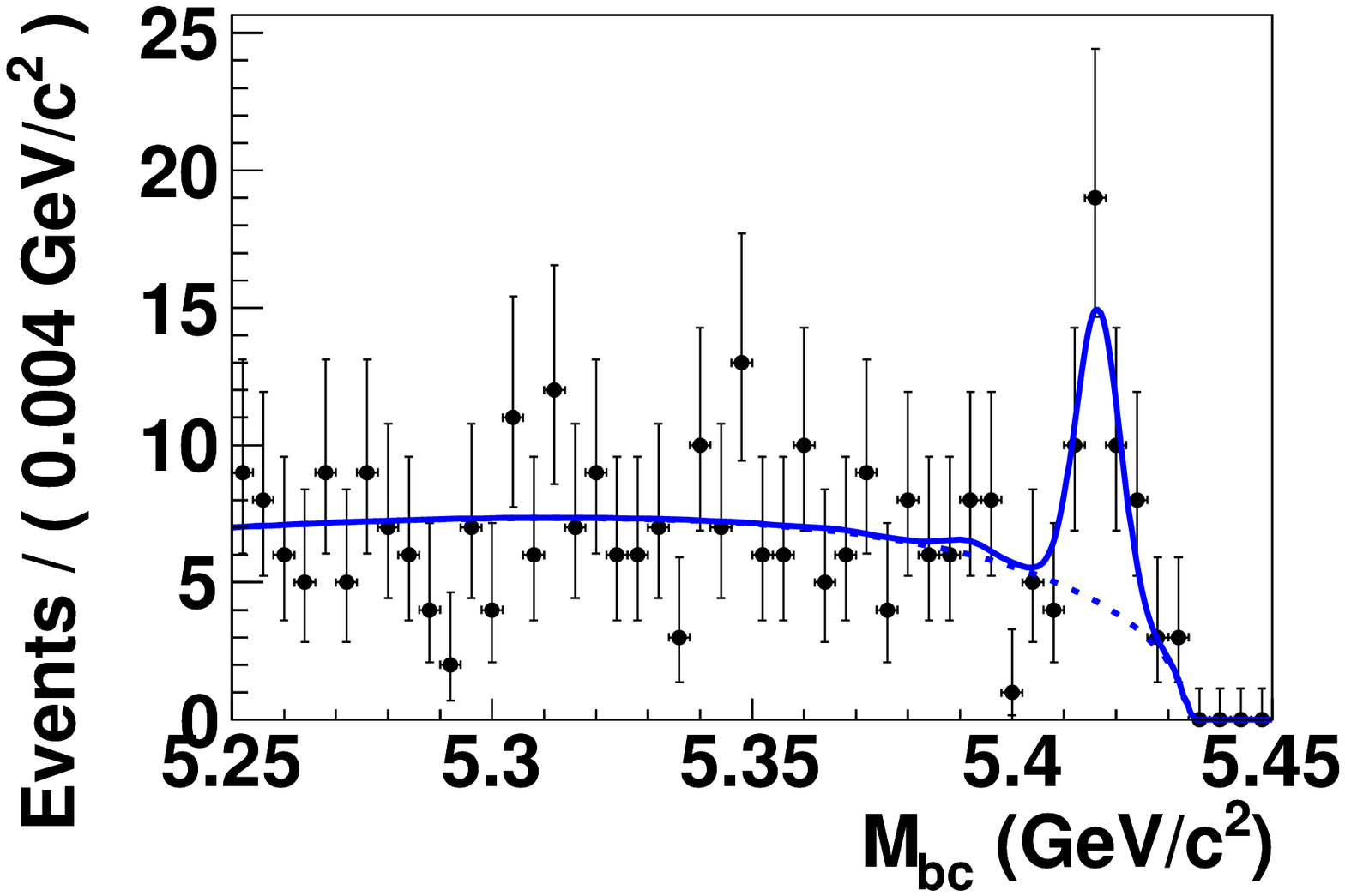}%
\hspace{-0.02\columnwidth}%
\includegraphics[width=0.53\columnwidth]
{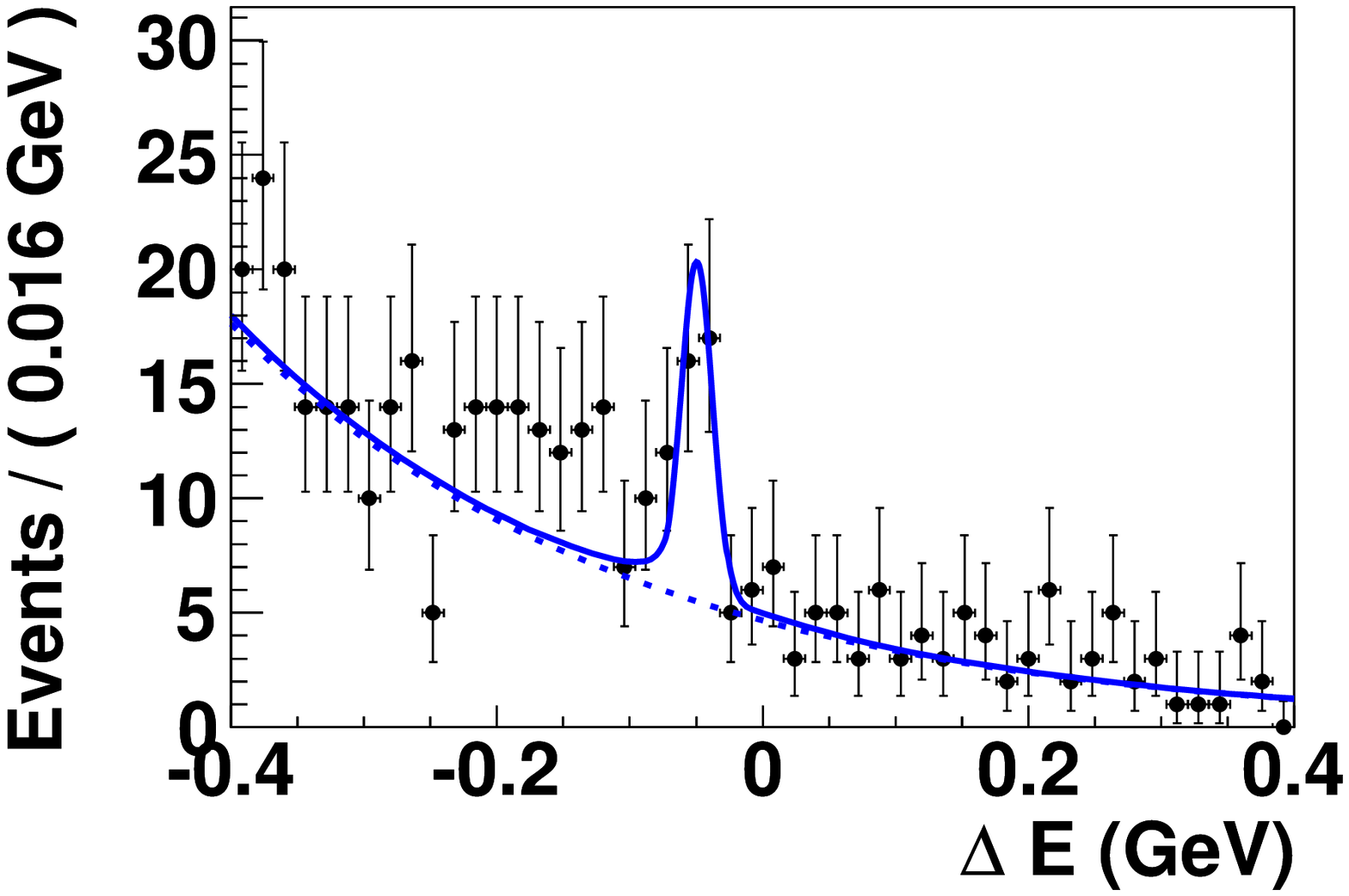}%
\hspace{-0.02\columnwidth}%
\hspace{-1\columnwidth}%
\begin{picture}(199,0)
\put(25,50){(c)}
\put(125,50){(d)}
\end{picture}\hfill}\\
\caption{\label{fig:5S_scan-epA-hmbc}
Fit projections for the clean $J/\psi\eta'(\eta\pi^+\pi^-)$ channel
with two $\eta$ subchannel combined (a), (b) and 
the $J/\psi\eta'(\rho^0\gamma)$ channel (c), (d). 
The projections are shown in the $B_s^*\bar B_s^*$ signal region with
$\Delta E\in [-87,-15]$ MeV (a), (c), and with
$M_\mathrm{bc}\in [5.405,5.429]\;\mathrm{GeV}/c^2$ (b), (d).
The sum of all backgrounds is represented by the blue dotted curves.}
\end{figure}

In the fit, the total probability density function consists of
a signal and background component. The signal component includes contributions
from the three $B_s^0$ pair production channels.  The signal normalization
for the $B_s^*\bar B_s^*$ production channel is parametrized as
$N_\mathrm{sig} = 2\times N_{B_s^{(*)}\bar B_s^{(*)}} f_{B_s^*\bar B_s^*}
\mathcal{B}(B_s^0\to J/\psi\eta^{(\prime)})
\mathcal{B}_i \epsilon_i $ for each $\eta^{(\prime)}$ subchannel $i$.
The product 
$\mathcal{B}_i= \mathcal{B}(J/\psi\to l^+l^-)\mathcal{B}_i(\eta^{(\prime)})$ is
the total branching fraction for a $J/\psi$ and  an $\eta^{(\prime)}$
decaying to the reconstructed final states~\cite{Nakamura:2010zzi},
and $\epsilon_i$ is the reconstruction efficiency obtained from
Monte Carlo simulation. The values of the weighted efficiencies
$\mathcal{B}_i\epsilon_i$ are listed in Table~\ref{tab:jAeA-N}.
The signal yields in the $B_s^*\bar B_s^0$ and  $B_s^0\bar B_s^0$
production channels are obtained in a similar manner, with
$f_{B_s^*\bar B_s^*}$ replaced by  $f_{B_s^*\bar B_s^0}$ and
$f_{B_s^0\bar B_s^0} = 1 - f_{B_s^*\bar B_s^*} - f_{B_s^*\bar B_s^0}$, respectively.
The floating parameters in the fit are the branching fractions
$\mathcal{B}(B_s^0\to J/\psi\eta^{(\prime)})$,
$f_{B_s^*\bar B_s^*}$, $f_{B_s^*\bar B_s^0}$, and the corresponding background
yields and shapes for different  $\eta^{(\prime)}$ subchannels.
This fit procedure was checked with six fully simulated Monte Carlo samples
that included both signal and background, each normalized to the data luminosity.
The results show that the fitted branching fractions for both modes recover
the input values.

The projections of the fit to the $121.4\;\mathrm{fb}^{-1}$ data sample
in the $B_s^*\bar B_s^*$ signal region are
shown in Figs.~\ref{fig:5S_scan-eA-hde_mbc} and
\ref{fig:5S_scan-epA-hmbc}.  There are good agreements between
fit curve and data points in all subchannels' projections.
We obtain a total of $141\pm 14$
$B_s^0\to J/\psi\eta$ events with a statistical significance of $21.9\sigma$
and $86\pm 14$ $B_s^0\to J/\psi\eta'$ events with a statistical significance
of $10.3\sigma$ in all three $\Upsilon(5S)\to B_s^{(*)}\bar B_s^{(*)}$ channels.
The statistical significances are calculated as 
$\sqrt{2\ln (L_\mathrm{max}/L_0)}$, where $L_\mathrm{max}$ and $L_0$ are the
maximum likelihood values, while the corresponding signal yield is set to zero for $L_0$.
The $B_s^0\to J/\psi\eta$ and $B_s^0\to J/\psi\eta'$ decays are observed
for the first time.  The $B_s^0$ pair production fractions are measured to be
$f_{B_s^*\bar B_s^*}=(90.5\pm 3.2\pm 0.1)\%$,
$f_{B_s^*\bar B_s^0}=(4.9\pm 2.5\pm 0.0)\%$, with a correlation coefficient $(-0.72)$.
This result is consistent with the value
 $f_{B_s^*\bar{B}_s^*} = (87.0 \pm 1.7)\%$~\cite{Li:2011pg}
obtained from $ 121.4\;\mathrm{fb}^{-1}$ of data using the $B_s^0 \to D_s^- \pi^+$
reconstruction method described in Ref.~\cite{:2008sc}.

\begin{table}
\caption{\label{tab:jAeA-N}
A summary of the product of the sub-branching fraction and efficiency
for various subchannels.  Here $\mathcal{B}_i= \mathcal{B}(J/\psi\to l^+l^-)
\mathcal{B}(\eta^{(\prime)}\to\mathrm{final\ state})$, with
the $J/\psi$ decaying to $e^+e^-$ or $\mu^+\mu^-$. }
\begin{ruledtabular}
\begin{tabular}{lc}
Subchannel & $\mathcal{B}_i\epsilon_i$  \\
\hline
$B_s^0\to J/\psi\eta(\gamma\gamma)$     & 1.40\%  \\
$B_s^0\to J/\psi\eta(\pi^+\pi^-\pi^0 )$ & 0.55\%   \\
Total $B_s^0\to J/\psi\eta$    & 1.95\% \\
$B_s^0\to J/\psi\eta'(\eta(\gamma\gamma)\pi^+\pi^-)$     & 0.45\%  \\
$B_s^0\to J/\psi\eta'(\eta(3\pi)\pi^+\pi^-)$     & 0.22\%  \\
$B_s^0\to J/\psi\eta'(\rho^0\gamma)$   & 0.96\%  \\
Total $B_s^0\to J/\psi\eta'$ & 1.63\% 
\end{tabular}
\end{ruledtabular}
\end{table}

\begin{table}
\caption{\label{tab:errBF_syst}
Relative systematic errors (in \%) for $\mathcal{B}(J/\psi\eta^{(\prime)})$.}
\begin{ruledtabular}
\begin{tabular}{lcc}
Source  &  $\mathcal{B}(J/\psi\eta)$ & $\mathcal{B}(J/\psi\eta')$\\ \hline
Signal shape calibration &  $+0.4, - 0.5$ &  $+1.1, -1.3$ \\
Track reconstruction  & 0.8 & 1.4  \\
Electron identification & 1.5 & 1.5\\
Muon  identification & 1.8 & 1.7\\
Pion identification   & 0.5  &  2.1  \\
$\eta(\pi^0)\to\gamma\gamma$ selection&  4.0 & 2.8  \\
$\mathcal{B}(J/\psi \to ll)$ & 0.7 & 0.7 \\
$\mathcal{B}(\eta^{(\prime)}\to \mathrm{final\ states})$ &
   0.5 & 1.2 \\
\hline
Total [without $N_{B_s^{(*)}\bar B_s^{(*)}}$] & 4.8 & 4.8  \\
$N_{B_s^{(*)}\bar B_s^{(*)}} $ & \multicolumn{2}{c}{$+22.4, -15.5$}
\end{tabular}
\end{ruledtabular}
\end{table}

The systematic uncertainties due to the signal function mean and width
are determined by varying each parameter by its error from the 
control sample calibration, repeating the fit,
and summing the shifts in the branching fraction in quadrature.
The lepton and pion identification efficiencies from Monte Carlo calculations
are calibrated
using $\gamma\gamma\to l^+l^-$ and $D^{*+}\to D^0\pi^+(D^0\to K^-\pi^+)$
control samples in data, respectively. 
Systematic errors for branching fractions are summarized in Table~\ref{tab:errBF_syst}. 
Those on $f_{B_s^*\bar B_s^*}$ and $f_{B_s^*\bar B_s^0}$ are dominated by
the signal shape uncertainty.
The large systematic error due to $N_{B_s^{(*)}\bar B_s^{(*)}}$
is quoted separately in the final results.

The ratio of the two branching fractions is also determined,
where the systematic error due to $N_{B_s^{(*)}\bar B_s^{(*)}}$ cancels.
For this, the statistical errors of the two modes are combined using
error propagation. Correlated systematic errors due to calibration, track
reconstruction, and particle identification are determined by
varying the numerator and denominator simultaneously.
Other systematic sources are treated independently.

In summary, we observe $B_s^0\to J/\psi\eta$
and  $B_s^0\to J/\psi\eta'$ decays for the first time with significances over
$10\sigma$, by taking advantage of the low background $e^+e^-$ environment at Belle.
We measure the branching fractions
\begin{align*}& \mathcal{B}(B_s^0\to J/\psi\eta)=\\ &\bsjetaBF, \\
&\mathcal{B}(B_s^0\to J/\psi\eta')=\\ &\bsjetapBF.
\end{align*}
These branching fractions are consistent with SU(3) expectations using the
measured value of $\mathcal{B}(B_d^0\to J/\psi K^0)$~\cite{Skands:2000ru}. 
The ratio of the two branching fractions is measured to be
$\frac{\mathcal{B}(B_s\to J/\psi\eta')}{\mathcal{B}(B_s\to J/\psi \eta)} =\bsrBF$.
This ratio is smaller than the expected value of $1.04\pm 0.04$
at the $2.1\sigma$ level; a significant deviation would indicate additional flavor
singlet components in the $\eta'$ other than $u\bar u,d\bar d,s\bar s$ pairs
or violation of the $\eta-\eta'$ mixing scheme.

We thank the KEKB group for excellent operation of the
accelerator, the KEK cryogenics group for efficient solenoid
operations, and the KEK computer group and
the NII for valuable computing and SINET4 network support.  
We acknowledge support from MEXT, JSPS and Nagoya's TLPRC (Japan);
ARC and DIISR (Australia); NSFC (China); MSMT (Czechia);
DST (India); INFN (Italy); MEST, NRF, NSDC of KISTI, and WCU (Korea); 
MNiSW (Poland); MES and RFAAE (Russia); ARRS (Slovenia); 
SNSF (Switzerland); NSC and MOE (Taiwan); and DOE and NSF (USA).
J. Li acknowledges support from WCU Grant No. R32-10155.

\end{document}